\documentclass[conference]{IEEEtran}
\IEEEoverridecommandlockouts
\usepackage{cite}
\usepackage{amsmath,amssymb,amsfonts}
\usepackage{algorithmic}
\usepackage{graphicx}
\usepackage{textcomp}
\usepackage{xcolor}
\usepackage{hyperref}
\usepackage{float}
\usepackage{subcaption}

\def\BibTeX{{\rm B\kern-.05em{\sc i\kern-.025em b}\kern-.08em
    T\kern-.1667em\lower.7ex\hbox{E}\kern-.125emX}}
\begin{document}

\title{Evaluating the Performance of Large Language Models in Competitive Programming: A Multi-Year, Multi-Grade Analysis\\
}

\author{
\IEEEauthorblockN{1\textsuperscript{st}Adrian Marius Dumitran}
\IEEEauthorblockA{\textit{Computer Science Department} \\
\textit{University of Bucharest}\\
Romania \\
marius.dumitran@unibuc.ro}
\and
\IEEEauthorblockN{2\textsuperscript{nd}Adrian Cătălin Badea}
\IEEEauthorblockA{\textit{Computer Science Department} \\
\textit{University of Bucharest}\\
Romania \\
adrian.badea15@unibuc.ro}
\and
\IEEEauthorblockN{3\textsuperscript{rd}Ștefan-Gabriel Muscalu}
\IEEEauthorblockA{\textit{Independent Researcher} \\
It Just Works Inc. \\
Bucharest, Romania \\
stefan.gabriel.muscalu@gmail.com}
}

\maketitle

\begin{abstract}
This study explores the performance of large language models (LLMs) in solving competitive programming problems from the Romanian Informatics Olympiad at the county level. Romania, a leading nation in computer science competitions, provides an ideal environment for evaluating LLM capabilities due to its rich history and stringent competition standards. We collected and analyzed a dataset comprising 304 challenges from 2002 to 2023, focusing on solutions written by LLMs in C++ and Python for these problems.

Our primary goal is to understand why LLMs perform well or poorly on different tasks. We evaluated various models, including closed-source models like GPT-4 and open-weight models such as CodeLlama and RoMistral, using a standardized process involving multiple attempts and feedback rounds. The analysis revealed significant variations in LLM performance across different grades and problem types. Notably, GPT-4 showed strong performance, indicating its potential use as an educational tool for middle school students. We also observed differences in code quality and style across various LLMs.
\end{abstract}


\section{Introduction}

This study evaluates the performance of various LLMs in solving competitive programming problems from the Romanian Informatics Olympiad at the county level (OJI - "Olimpiada Județeană de Informatică"). Our primary objective is to understand why LLMs excel or struggle with specific tasks. We analyze data spanning over two decades, focusing on challenges written in C++, the predominant language used in programming competitions, and Python, the language usually used in papers that test LLM code generation.

We collected and cleaned data from multiple sources, including historical archives and recent competition records. The dataset comprises 304 computer science challenges from 2002 to 2023, written in Romanian.

Our findings reveal significant variations in LLM performance across different grades and problem types. GPT-4, for instance, demonstrated strong performance in lower grades but struggled with more complex problems typically encountered in higher grades. We also observed notable differences in code quality, with GPT-4 generating more verbose and production-ready code compared to other models, which produced more concise and straightforward solutions.

This research provides valuable insights into the capabilities and limitations of LLMs in competitive programming contexts. By understanding the factors that influence LLM performance, we can better design models and training datasets to enhance their problem-solving abilities. Furthermore, these findings have practical implications for educational settings, suggesting ways to leverage LLMs to support personalized learning and improve competitive programming training.

\subsection{About OJI, county level Olympiads in Romanian}

Romania is one of the top performers in computer science competitions for middle and high school students, placing 3rd in the all-time ranking of countries\footnote{\url{https://stats.ioinformatics.org/countries/?sort=total_desc}}. With such a strong history, the national Olympiad generates a lot of interest, but out of tens of thousands of participants, only about 600 qualify for the national phase.

The final stage of qualifying is the OJI ('Olimpiada Județeană de Informatică'), where participants compete within their county on common subjects.

These participants are organized into eight distinct classes. Approximately 80 students from each class advance to the nationals, with the top performer from each county guaranteed a spot by default.

In terms of problem-solving, middle school students tackle two problems, while high school students face between two and three problems, particularly after the year 2017 when this format was standardized.

Competition problems are exclusively designed in C++, the predominant language used by most contestants. Although Pascal is permitted within the rules of the competition, its usage among competitors is minimal.

Participants are allocated between three to four hours to solve these challenges during an in-person event.

Scoring in this competition allows for partial credit; some problems may include anywhere from ten to thirty tests designed to evaluate various aspects of algorithmic efficiency and correctness under different scenarios. This scoring system ensures that even partially correct solutions receive recognition towards a participant's total score.

\section{Related Work}
The capabilities of large language models (LLMs) for code generation have been extensively explored in the literature, particularly in the context of competitive programming contests. Authors from \cite{hendrycks_et_al} were pioneers in this area, creating a diverse dataset of 10,000 programming problems that span various difficulty levels. This dataset has been foundational for subsequent research in the field. Another notable contribution is the TACO dataset \cite{li_rongao_et_al}, which offers a broad array of programming contest problems and benchmarks the performance of various LLMs in code generation and program tagging. Additionally, AlphaCode \cite{li_yujia} represents a significant development, providing an extensive dataset that has markedly enhanced the training of LLMs for competition-level code generation.

Among these resources, RoCode \cite{cosma_et_al} stands out as it is the only dataset that includes problems in Romanian, offering a unique perspective on assessing code intelligence. Despite the availability of rich datasets and benchmarks, current research predominantly focuses on quantitative analyses and often overlooks qualitative aspects, such as the nuanced strategies employed in problem-solving within competitive programming. 

Our "OJI" dataset is unique in that it incorporates problems from the same competition, with a defined curriculum for each class, allowing for a more fine-grained analysis.

A more focused dataset, HumanEval \cite{chen_et_al}, encompasses a smaller yet insightful collection of problems used to evaluate the capabilities of code-centric LLMs. This dataset has been instrumental in refining evaluation techniques, which we have adopted and expanded upon in our study.

Significantly, existing literature does not adequately address code generation across multiple programming languages, nor does it specifically focus on C++, despite its prevalence in competitive programming contests. Our research addresses this gap by evaluating LLM performance in both Python and C++, with a particular emphasis on C++. This approach is supported by our newly developed evaluation system, designed to comprehensively assess the efficacy of LLMs in a multilingual, competitive programming context.

\section{Methodology}

\subsection{Data Collection}

Data was collected from \href{https://kilonova.ro/problem_lists/460}{Kilonova}.
The dataset comprises 304 computer science challenges from the 2002-2023 editions of the Romanian Computing Olympiad, at the county stage, written in Romanian. 

Results were gathered from multiple sources to ensure a comprehensive dataset. Historical data from 2002 to 2020 was obtained through the archived records available on \href{https://olimpiada.info/}{olimpiada.info}, despite some gaps and incomplete entries. For the period from 2021 onwards, we collected results from \href{https://sepi.ro/}{sepi.ro}, which provided more complete and up-to-date records.

\subsection{Data Cleaning}

There were several challenges in normalizing the data collected from Kilonova into a format suitable for analysis. A parser was developed to extract the data into different sections such as metadata, year, grade, statement, input, output, constraints, and examples. 
The results gathered from olimpiada.info and sepi.ro were parsed and aggregated manually.  

\subsection{Choosing the LLMs to Evaluate}

The LLMs selected for evaluation include both closed and open-source models. The models evaluated were of the \textit{chat}/\textit{instruct} type and are as follows:
\begin{itemize}
    \item GPT-4 1106 - Limited API access
    \item Gemini 1.0 Pro - Provided freely by Google during beta testing
    \item Codestral (22B)
    \item DeepSeek Coder (6.7B, 33B)\cite{deepseek-coder}
    \item AutoCoder (6.7B, 33B)\cite{autocoder}
    \item CodeLlama (7B, 13B)\cite{codellama}
    \item Phind-CodeLlama (34B)
    \item CodeQwen 1.5 (7B)
    \item Mistral (7B)\cite{mistral}
    \item Llama 3 (8B)
    \item Yi (9B)
    \item Phi3 (14B)\cite{phi3}
    \item StarCoder 2 (15B)\cite{starcoder2}
    \item RoLlama2 (7B)\cite{openllm-ro}
    \item RoMistral (7B)\cite{openllm-ro}
\end{itemize}

The closed-source models were chosen for their availability and ease of use, while the open-weight models were selected for their perceived strong performance in programming tasks and being below 33B.

\subsection{Evaluation Methodology}

All LLMs were given the same structure when prompting and providing feedback. Each model was given multiple attempts (\(k=5\) for closed-source, \(k=3\) for open-weights) to solve each problem. Each attempt allowed a maximum of feedback rounds (\(f=5\) for closed-source, \(f=3\) for open-weights) to complete the challenge. Closed-source models were tasked with solving the entire dataset, while open-weight models were only assigned grade 5 challenges. Unless specified, the best run out of all attempts per challenge is the only one being taken into account when creating statistics. 
Overall, around 22,700 attempts were computed on the full dataset using closed-source models, and around 3,280 attempts were computed on the open-weights models for the grade 5 subset of the dataset. 

\subsection{Conversation and Prompting}

\begin{enumerate}
    \item The initial prompt consisted of a brief description setting up the expectations from the model (programming language, input/output file names, or console interaction), followed by the problem statement, input, output, and one example.
    \item The model was then expected to respond with a solution to the challenge. If the model failed to provide a solution (early stop), it was re-prompted to continue from where it left off.
    \item The solution was compiled and evaluated in a Docker container using the \textit{gcc:11} image\footnote{\href{https://hub.docker.com/layers/library/gcc/11/images/sha256-97f939499d822bfda05e5398379cfe78d0b903a154201e9ff56da9358597a356?context=explore}{GCC:11 Dockerhub}} in the C++ tests, and evaluated using the Docker \textit{python:3.11} image \footnote{\href{https://hub.docker.com/layers/library/python/3.11/images/sha256-487f28cb57a7a1a1a0a40bed065850fd7ed1c11cd1acd5dfcbb6aa0e05994fc9?context=explore}{Python:3.11 Dockerhub}} for python tests. During this evaluation, all examples were tested and results were compared against the expected outcome.
    \item[4a)] If the solution was correct, the code was recorded and the process moved to step 5.
    \item[4b)] If the solution was incorrect, the model received feedback on the error (compilation error, runtime error, wrong answer) and was re-prompted to fix the mistake and rewrite the solution. Steps 2 through 4 were repeated until the feedback rounds were exhausted or the solution was correct.
    \item[5)] When the solution passed the examples or the feedback rounds were exhausted, the solution, if correct, was submitted to \href{https://kilonova.ro/}{Kilonova} for a comprehensive evaluation and grading.
\end{enumerate}

\begin{figure}[H]
\centering
\includegraphics[width=1\linewidth]{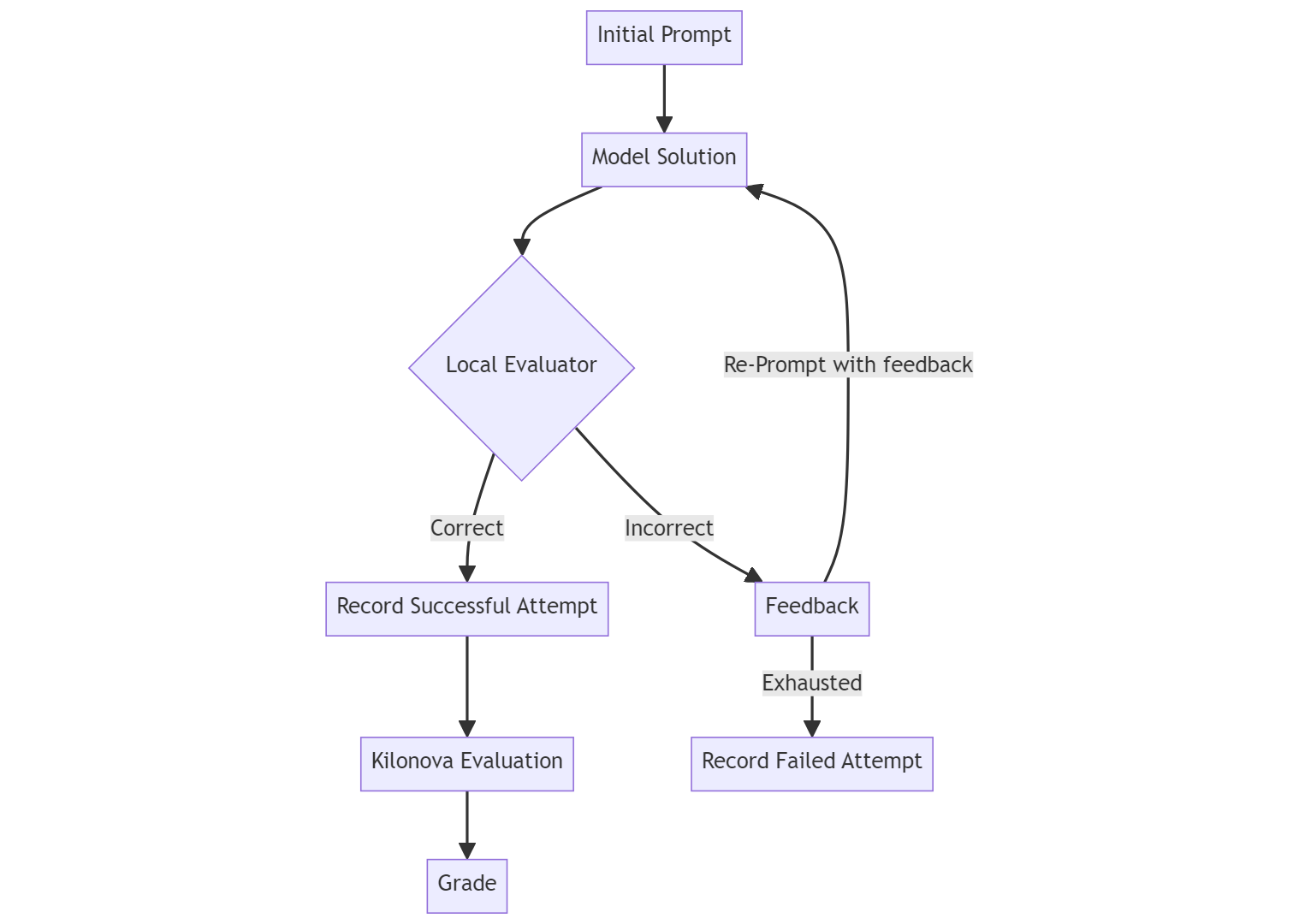}
\caption{Evaluation Flowchart}
\label{fig:methodology_flowchart}
\end{figure}

\subsection{Running the LLMs}

All conversations were conducted using the model's conversational API. For closed-source models, the API was used to send and receive responses. For open-weight models, their GGUF variants were used in conjunction with an Ollama server to handle the conversations. Tests were conducted on an RTX 4090 GPU. Models below 10B parameters were run in FP16 format, those below 20B in quantized Q8\_0, and those above 20B in quantized Q6\_K / Q4\_K formats, described as in \cite{quantization}\cite{quantization2}.


There were around 112,400 submissions which initially underwent a series of local evaluations. These are preliminary checks done at a smaller scale or earlier stage to ensure that each submission meets certain criteria or standards. After passing these initial evaluations, the submissions then proceeded to be evaluated by the Kilonova evaluator.

The average attempt took around 01m:56s, with duration as low as 00m:16s (at p05) and as high as 04m:28s (at p97), totaling all closed-source and open-weight models execution time to 31 days, 17h:05m:40s.

\section{Results}

\begin{figure}[H]
    \centering
    \includegraphics[width=1\linewidth]{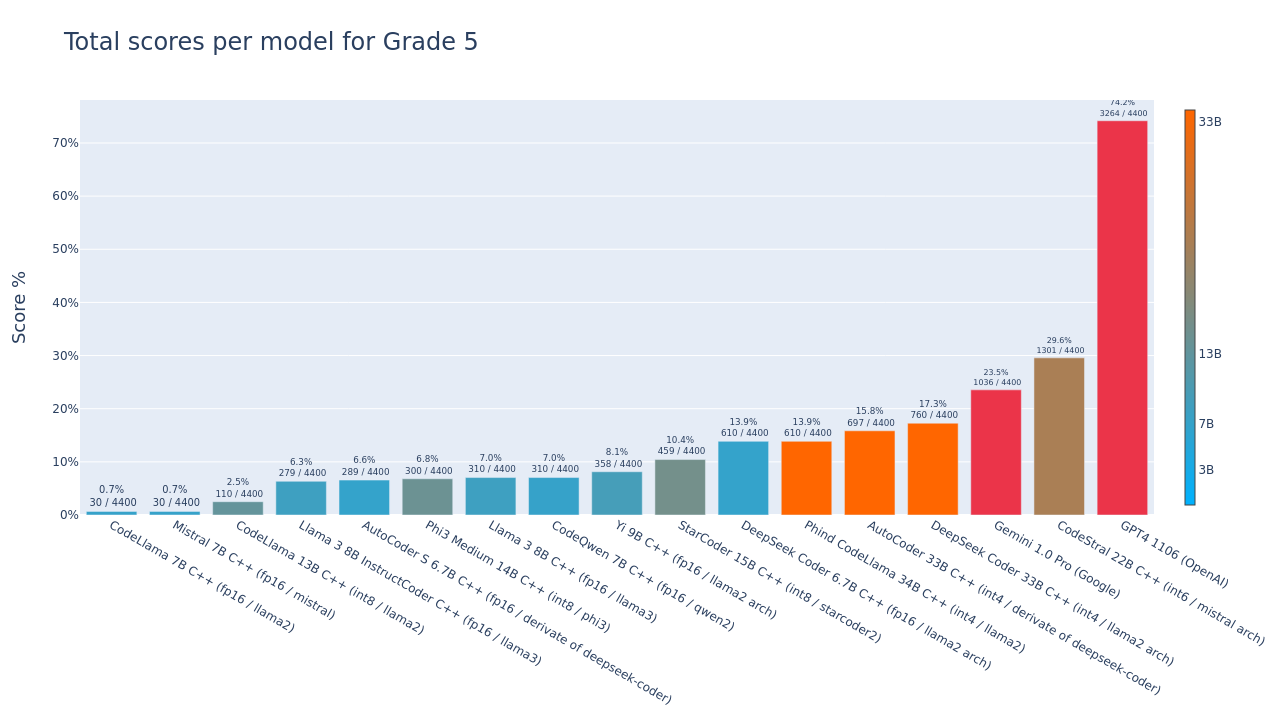}
    \caption{Total Scores in Grade 5 per Model}
    \label{fig:grade-5-scores}
\end{figure}

\subsection{LLM Comparison}

\subsubsection{Open-Weights Small Models}
As illustrated in the figure ~\ref{fig:grade-5-scores}, there is a noticeable trend where smaller models exhibit reduced performance in solving the challenges. Even models with parameters as large as 33B struggle to match the performance of Gemini 1.0 Pro, with the exception of one of the latest models, Codestral 22B\footnote{\url{https://huggingface.co/mistralai/Codestral-22B-v0.1}}. This performance discrepancy is somewhat expected given the nature of the challenges difficulty and the limited training most models have with Romanian text.

We also experimented with RoLlama2 7B\footnote{\url{https://huggingface.co/OpenLLM-Ro/RoLlama2-7b-Chat}}\cite{openllm-ro} and RoMistral 7B\footnote{\url{https://huggingface.co/OpenLLM-Ro/RoMistral-7b-Instruct}}\cite{openllm-ro}, which are fine-tuned versions of Llama 2 and Mistral models specifically on Romanian text. However, their performance was significantly degraded compared to their base models, as they failed to generate correct code syntax in English.

\subsubsection{Code Quality and Particularities}

\begin{figure}[H]
\centering
\begin{subfigure}{.155\textwidth}
    \centering    \includegraphics[width=1.2\linewidth]{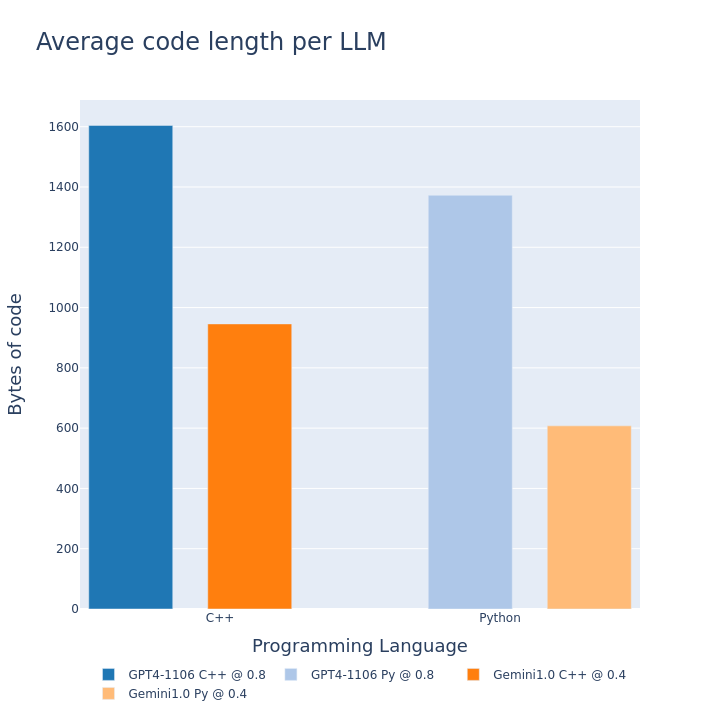}  
    \caption{Avg File Size}
    \label{code_bytes}
\end{subfigure}
\begin{subfigure}{.155\textwidth}
    \centering
\includegraphics[width=1.2\linewidth]{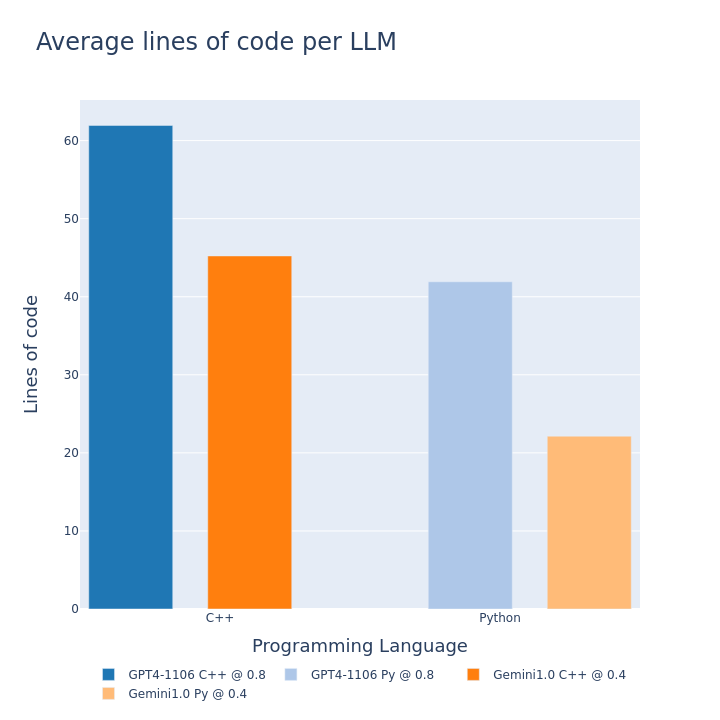}
    \caption{Avg Line Count}
    \label{code_lines}
\end{subfigure}
\begin{subfigure}{.155\textwidth}
    \centering
    \includegraphics[width=1.2\linewidth]{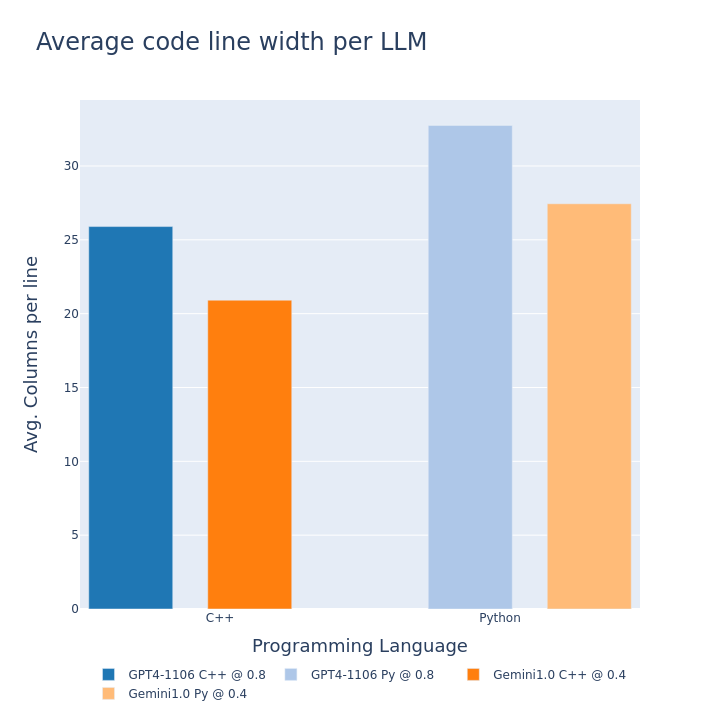}  
    \caption{Avg Line Width}
    \label{code_width}
\end{subfigure}

\caption{Code Size}
\label{fig:CodeLength}
\end{figure}

Inspired by research in \cite{cosma_et_al}, we examined the length of the code generated by LLMs. While the difference between the length of correct and incorrect solutions was not significant (approximately 10\%), we observed a notable disparity in the code length generated by GPT-4 compared to Gemini 1.0 PRO, as shown in Figure \ref{fig:CodeLength}. As seen in the figure GPT-4 not only has more lines but also has longer lines. Variable names are one of the causes for this, for example Gemini uses h, m, s whereas GPT4.0 uses structures and uses startTime.h, starTime.m, startTime.s.

\begin{figure}[H]
    \centering
    \includegraphics[width=0.45\textwidth]{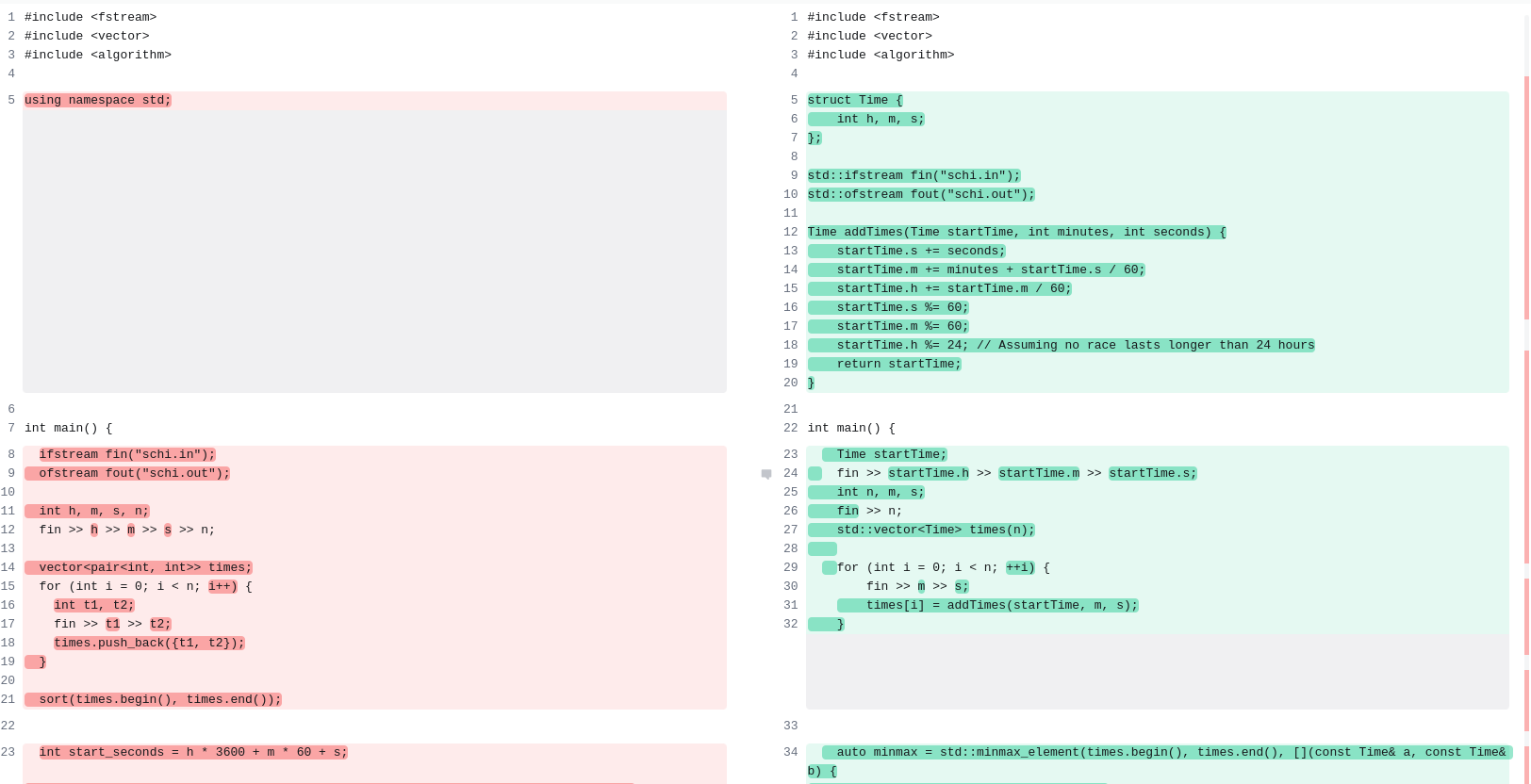}
    \caption{Code Length per LLM}
    \label{fig:CodeDiff}
\end{figure}

This significant difference prompted further investigation. We found that Gemini tends to produce simpler code, similar to what might be expected from a novice programmer. In contrast, GPT-4 generates more comprehensive, production-ready code, which includes more comments, advanced features such as functions and structures, and sophisticated language constructs like `std::find`. Such differences can be seen in figure\ref{fig:CodeDiff}

Although such detailed and well-structured code is advantageous in a production environment, competitive programming often favors more concise and efficient code.

We have also investigated why ChatGPT-4 sometimes fails to solve certain problems and discovered that it struggles with tasks that involve multiple components, such as this problem \footnote{\url{https://kilonova.ro/problems/512}}. It can easily solve case 1 but struggles with cases 2 and 3. Even when provided with assistance, it may solve case 2 but then forget to address case 1, and so on.

This raises the question of whether writing extensive code might impede GPT-4's ability to solve tasks with more complex structures, even if the tasks themselves are not inherently difficult.

\subsection{GPT4 contestant and teacher}

\begin{table}[H]
\centering
\begin{tabular}{|l|c|c|c|c|}
\hline
\textbf{name} & \textbf{Grade} & \textbf{No. Qualified} & \textbf{Total} & \textbf{Percentage} \\
\hline
GPT4-1106 C++ & 5  & 8 & 8 & 100.0\% \\
              & 6  & 4 & 8 & 50.0\% \\
              & 7  & 5 & 8 & 62.5\% \\
              & 8  & 4 & 7 & 57.1\% \\
              & 9  & 1 & 9 & 11.1\% \\
              & 10 & 6 & 10 & 60.0\% \\
              & 11 & 2 & 7 & 28.6\% \\
              & 12 & 2 & 8 & 25.0\% \\
\hline
\end{tabular}
\caption{GPT4 Qualifications for National Olympiad}
\label{tab:GPT4Competitive}
\end{table}

Table \ref{tab:GPT4Competitive} shows GPT-4 C++ performance compared to pupils in county-level Olympiads. Despite GDPR challenges, we collected 63 data points, with at least 7 per grade. GPT-4 qualified in 100\% of 5th grade cases and over 50\% for 6th, 7th, 8th, and 10th grades but struggled with 9th, 11th, and 12th grades due to problem complexity for 11th and 12th grades and higher competition in 9th grade.

Given the impressive results of GPT-4, there are numerous opportunities for its use in educational settings:

\textbf{Personalized Learning:}
\begin{itemize}
    \item \textbf{Strength Identification:} GPT-4 can pinpoint students' strengths and weaknesses through detailed feedback.
    \item \textbf{Targeted Practice:} Generates specific practice problems to address weak areas.
\end{itemize}

\textbf{Immediate Feedback:}
\begin{itemize}
    \item \textbf{Real-time Assistance:} Provides instant help with problem-solving and concept understanding.
    \item \textbf{Solution Explanation:} Offers step-by-step explanations to enhance comprehension.
\end{itemize}

Furthermore, LLMs have the potential to generate programming contests. However, this introduces several challenges:

\begin{itemize}
    \item \textbf{Fairness and Integrity:} Given that LLMs can solve many problems at this level, there is a risk of unfair use by students. Strict monitoring and innovative problem design are necessary to address this.
    \item \textbf{Problem Complexity:} Creating problems that LLMs cannot yet solve or modifying tasks to require human-LLM collaboration could be effective strategies.
    \item \textbf{Ethical Considerations:} The use of LLMs in contests should be managed to ensure that they are used to enhance learning rather than undermine the competition's integrity.
\end{itemize}

Balancing the benefits of GPT-4 in educational settings with the need to maintain fair and challenging contests is crucial for fostering a healthy and effective learning environment.

\subsection{Year to year comparison}
Our dataset includes OJI problems spanning over more than 20 years. We focused our analysis on ChatGPT (though Gemini and CodeStral for 5th grade results appear in the f\ref{fig:performance_by_class}), which was the only LLM to achieve competitive results. Through this analysis, we identified several inflection points where the difficulty of problems increased, even though these increases were not continuous. Surprisingly, while the curriculum has not changed significantly over the years, we observed that the difficulty of the problems increased over time until 2020, punctuated by several notable inflection points::

\begin{figure}[H]
    \centering
    \includegraphics[width=0.45\textwidth]{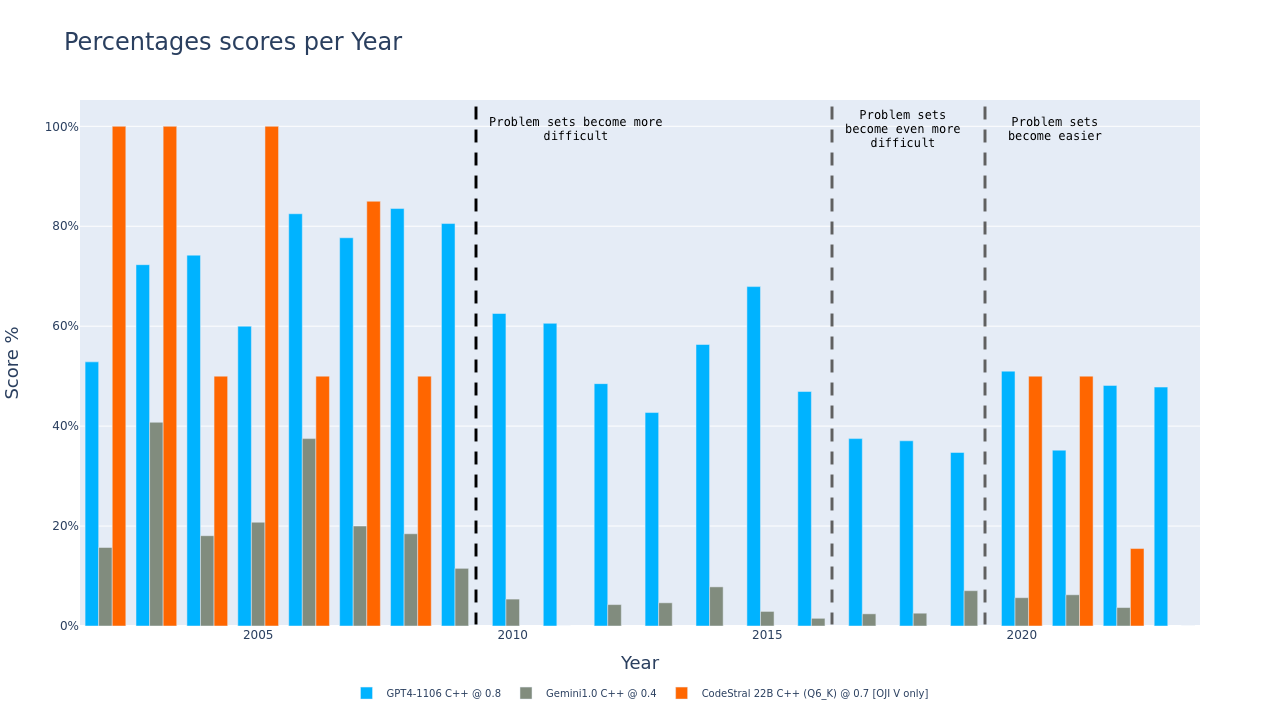}
    \caption{Trends in OJI Problem Difficulty Over 20 Years}
    \label{fig:oji_by_year}
\end{figure}

\begin{enumerate}
    \item \textbf{2010:} The National Committee decided to introduce more difficult problems to enhance competitiveness for the IOI.
    \item \textbf{2017:} The OJI High School competition began featuring three problems, typically consisting of two difficult problems and one easy problem, rather than the previous format of one easy and one hard problem. This change increased the overall average difficulty.
    \item \textbf{2020:} Qualifying for the national competition became easier, prompting the National Committee to decrease the average difficulty of the problems.
\end{enumerate}

\subsection{Grade by Grade}

\begin{figure}[H]
    \centering
    \includegraphics[width=0.45\textwidth]{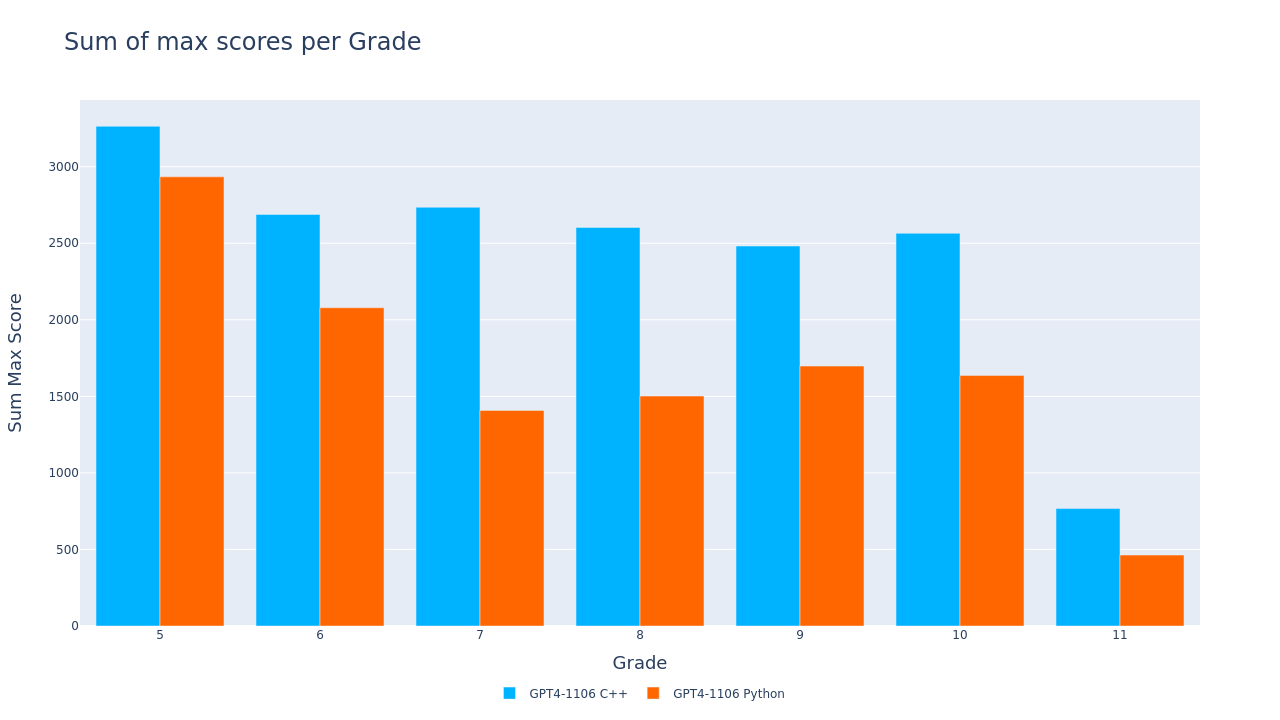}
    \caption{Score Percentage per Grade}
    \label{fig:performance_by_class}
\end{figure}

The graph provided (Figure \ref{fig:performance_by_class}) shows the LLM's performance in solving CS problems for grades 5 through 12, coded in C++ and Python. We analyze the curriculum\footnote{\url{https://sepi.ro/assets/upload-file/oni2024/Programa\%20pentru\%20olimpiada\%20de\%20informatica\_gimnaziu\%20si\%20liceu.pdf}} content to understand why certain performance patterns emerge.

\begin{itemize}
    \item \textbf{Grade 5:} 
    \begin{itemize}
        \item \textbf{Content:} Basic algorithms, simple data types, control structures, number processing.
        \item \textbf{Performance:} High (74\%). The curriculum involves fundamental concepts, making it easier for the LLM.
    \end{itemize}
    
    \item \textbf{Grade 6:} 
    \begin{itemize}
        \item \textbf{Content:} Real numbers, character types, modular arithmetic, prime factorization.
        \item \textbf{Performance:} Drop to 61\%. Increased complexity with abstract concepts. LLM's struggle with solving mathematical problems.
    \end{itemize}
    
    \item \textbf{Grade 7:} 
    \begin{itemize}
        \item \textbf{Content:} Functions, advanced array techniques, greedy methods.
        \item \textbf{Performance:} Increase to 62\%. Systematic approach and modular problems align with LLM's capabilities.
    \end{itemize}
    
    \item \textbf{Grade 8:} 
    \begin{itemize}
        \item \textbf{Content:} Character strings, combinatorial algorithms, basic queue operations.
        \item \textbf{Performance:} High (59\%). Proficiency in handling string operations and combinatorial problems.
    \end{itemize}
    
    \item \textbf{Grade 9:} 
    \begin{itemize}
        \item \textbf{Content:} Binary search, prefix sums, greedy methods.
        \item \textbf{Performance:} Drop to 52\%. Complexity of algorithms and data structures increases significantly.
    \end{itemize}
    
    \item \textbf{Grade 10:} 
    \begin{itemize}
        \item \textbf{Content:} Stack, queue, deque operations, advanced data structures.
        \item \textbf{Performance:} Small drop to 50\%. Focus on data structures introduces intricate problems.
    \end{itemize}
    
    \item \textbf{Grade 11-12:} 
    \begin{itemize}
        \item \textbf{Content:} Dynamic programming, complex graph algorithms, advanced tree structures.
        \item \textbf{Performance:} Significant drop to 26\%. Challenges with deep abstraction and sophisticated problem-solving.
    \end{itemize}
\end{itemize}

The LLM's performance aligns with the increasing complexity of the curriculum. Higher proficiency is observed with simpler, structured problems, while significant challenges arise with advanced topics requiring deep abstraction and intricate algorithms. Consistent performance from grades 7 to 10 indicates the LLM's ability to manage complex but pattern-recognizable problems, whereas a sharp decline at grade 11 highlights its difficulty with dynamic programming and graph theory.

\subsection{Comparison of C++ and Python Results}

This subsection is dedicated to contrasting the efficiency of two programming languages, C++ and Python, in the context of the GPT code generation test. The outcomes revealed substantial disparities in the efficiency of the two languages. The analysis focuses on languages that were accessible for both models: GPT-4-1106-Preview and Gemini Pro 1.0.

As depicted in Figure \ref{fig:performance_by_class} and  \ref{fig:oji_by_year}, C++ outperformed Python for all difficulties.

\textbf{C++ advantages over Python}
\begin{itemize}
    \item The committee's preference for C++ as the official source language might one of the reasons for this. This preference is not arbitrary but is based on several factors, including better management of soft and hard resources. 
    \item An additional factor could be the limited language options available to competitors, which are confined to C/C++ and Pascal.
    \item Furthermore, C++ has a robust standard library that includes a rich set of functions, algorithms and data structures, making it a versatile language for various programming tasks. Data structures also exist in Python, but they are not so popular and probably do not appear in the majority of training code. 
\end{itemize}

As shown in Figure \ref{fig:performance_by_class}, the performance of C++ was better than that of Python for GPT-4-1106-Preview.

\begin{figure}[h]
    \centering
    \includegraphics[width=0.45\textwidth]{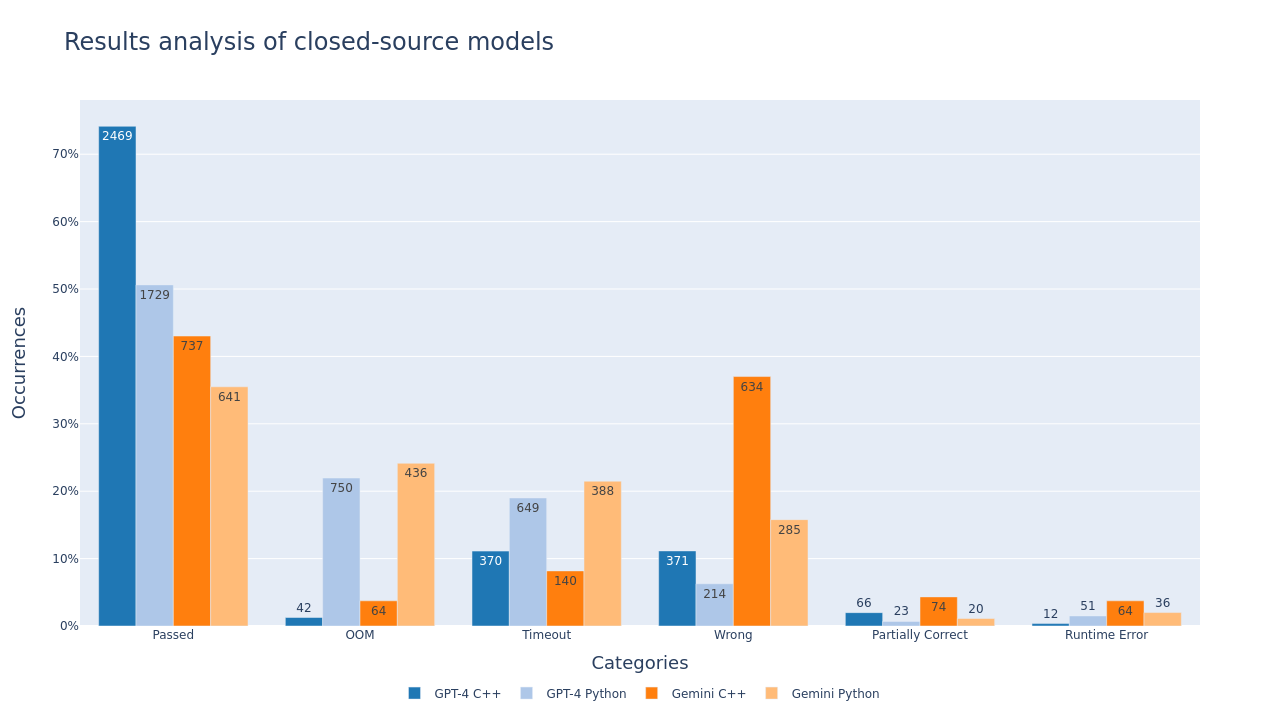}
    \caption{Result categories of closed-source models}
    \label{fig:performance_comparison}
\end{figure}

That being said, there are specific tasks that Python does handle better. Python has an advantage in solving problems related to strings and to big integers. For example, in the following problem\footnote{\url{https://kilonova.ro/problems/794}}, Python obtains 100 points, whereas C++ only 35 as it tries to transform a 2000-digit string to an integer (\texttt{stoll(num\_combined);}\ldots which is possible in Python but not in C++).

Another example is this problem \footnote{\url{https://kilonova.ro/problems/928}}, a classical \textbf{strings} problem. For this problem, Python obtains 100 points whereas C++ obtains 0.

In our code generation example, Python encountered issues with memory exhaustion and time constraints, as shown in figure \ref{fig:performance_comparison}

It's evident that Python's primary application is not competitive programming, as it is typically utilized in environments without strict boundaries, such as training models and artificial intelligence applications.

\subsection{Temperatures}

\begin{figure}[h]
    \centering
    \includegraphics[width=0.45\textwidth]{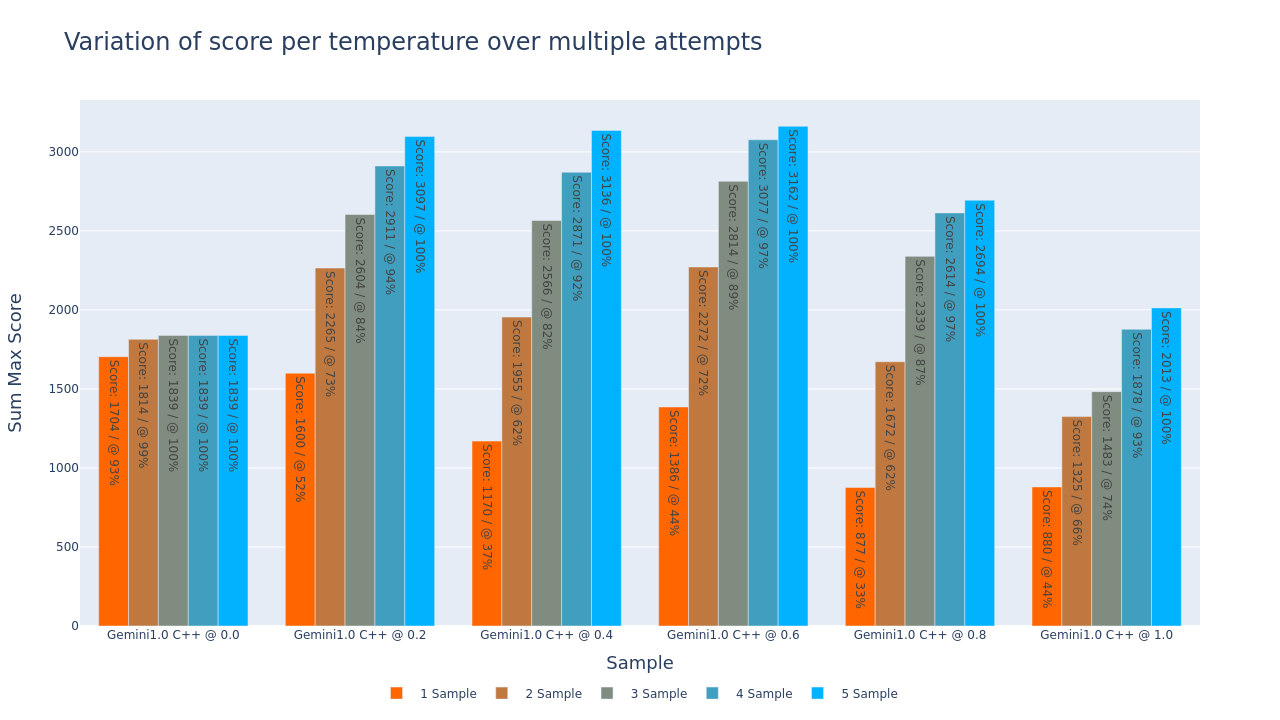}
    \caption{Variation of scores on temperatures over attempts}
    \label{fig:variation_score_per_temp_over_attempts}
\end{figure}

\textbf{Temperature Analysis}
\begin{itemize}
    \item At temperature 0.0, scores remain relatively consistent across sample sizes, with scores ranging from 1704 to 1839.
    
    \item At temperature 0.2, there is a notable increase in scores as the number of samples increases, peaking at 2911 with 4 attempts.
    
    \item For temperatures 0.4 and 0.6, scores increase significantly with the number of samples, with the highest scores being 3136 and 3162 respectively for 5 samples.
    
    \item At temperature 0.8, the scores plateau and peak at 2694 for 5 samples.
    
    \item At temperature 1.0, the scores show a similar pattern, with a peak score of 2013 for 5 samples, indicating a slight decline compared to 0.8.
    
    \item Comparable to findings in previous studies \cite{renze2024effect}, problem-solving performance starts to notably deteriorate when the temperature surpasses 1.0.
    
\end{itemize}

\textbf{Sample Size Impact}
\begin{itemize}
    \item Single sample scores are consistently lower across all temperatures.
    
    \item A significant improvement is seen when increasing the sample size from 1 to 3 samples, particularly at mid to high temperatures (0.4 to 0.8).
    
    \item The largest gains from increasing sample size are observed at temperatures 0.6 and 0.8, where scores almost double from 1 sample to 5 samples.
\end{itemize}

\paragraph{Score and Completion Percentage}
\begin{itemize}

    \item The score and completion percentages are annotated on the bars. Higher scores generally correlate with higher completion percentages.
    
    \item Completion percentage tends to be higher at higher sample sizes and higher temperatures, indicating more reliable performance under these conditions.
\end{itemize}

\section {Conclusions}
\subsection{Findings}
\begin{enumerate}
    \item Our study demonstrates that LLMs are more efficient in solving Olympiad problems when generating C++ code. Most existing studies have focused on generating Python code, the most frequently used by LLMs. However, our findings show that C++ performs better for these tasks.
    \item While LLMs can be competitive in county level olympiads, researches such as \cite{huang_et_al} emphasize the critical need for robust datasets in order to accurately evaluate LLMs’ reasoning abilities.  
    
    \item ChatGPT-4 generates more verbose and production-ready code, while Gemini 1.0PRO produces more concise and straightforward code, typical of what is expected in programming competitions.
    
    \item Small open-weight models are good for tasks such as "fill in the middle" but struggle to write a correct solution from start to finish.
    
    \item New open-weight models are emerging that demonstrate better capabilities than closed-source counterparts. 
    
\end{enumerate}
\subsection{Limitations}
\begin{enumerate}
    \item Open-weight models were constrained by their maximum context size, which was set according to the model's specifications. This limit was often reached by smaller models, causing evaluation to stop prematurely.
    \item Open-weight models were constrained by our testing environment. All models above 10 billion parameters were quantized to Q8, Q6 or Q4\cite{quantization}\cite{quantization2} in order to be evaluated in a feasible time-frame. 
\end{enumerate}

\section{Further Work}

\subsection{Tagging}
Ongoing efforts are focused on tagging each challenge based on the type of optimal solution required to achieve the maximum score, as well as identifying any possible partial solutions tags. \cite{tagging}

\subsection{Difficulty Assessment}
The current study assumes that difficulty increases with grade level. While this is generally accurate, a detailed expert assessment of all challenges would be beneficial for determining their true difficulty levels.

\subsection{Translation to English}

Further work includes translating the dataset into English to compare open-weight and closed-source models. This aims to identify score losses due to misunderstandings of the challenge statements. However, naive translations may lose important nuances.

\subsection{Human + LLM}
While LLMs can solve many problems independently, a human-LLM collaboration is expected to be even more effective. We aim to experiment with various settings to explore how humans and LLMs can work together to solve problems more efficiently.

\subsection{ChatGPT Augmented Challenges}
We are exploring the potential of combining GPT models with human experts in the creation of new challenges for future contests. This approach aims to enhance the quality and diversity of the challenges.

\end{document}